\documentclass[11pt, a4paper]{article}
\usepackage[affil-it]{authblk}

\usepackage[sort,nocompress]{cite}
\usepackage[utf8]{inputenc}
\usepackage{tabularx}
\usepackage{xcolor}
\usepackage{graphicx}
\usepackage{multicol}
\usepackage{varwidth}
\usepackage{placeins}
\usepackage{hyperref}
\usepackage{amsfonts}
\usepackage{amsmath}
\usepackage{booktabs}
\usepackage{multirow}
\usepackage{enumitem}
\usepackage{subfiles}
\usepackage{caption}
\usepackage{subcaption}

\newcommand{\PLH}{{\mkern-2mu\times\mkern-2mu}}
\graphicspath{{images/}}

\begin{document}

\title{SpineNetV2: Automated Detection, Labelling and Radiological Grading Of Clinical MR Scans}
\author[1]{Rhydian Windsor}
\author[1]{Amir Jamaludin}
\author[1,2]{Timor Kadir}
\author[1]{Andrew Zisserman}
\affil[1]{Visual Geometry Group, 
       Department of Engineering Science, 
       University of Oxford}
\affil[2]{Plexalis}
\affil[ ]{\normalfont{\texttt{\{rhydian,amirj\}@robots.ox.ac.uk}}}
\maketitle
\begin{abstract}
This technical paper presents SpineNetV2, an automated tool which:
(i) detects and labels vertebral bodies in clinical spinal magnetic 
resonance (MR) scans across a range of commonly used sequences;
and (ii) performs radiological grading of lumbar intervertebral discs in T2-weighted
scans for a range of common degenerative changes.
SpineNetV2 improves over the original SpineNet software in two ways:
(1) The vertebral body detection stage is sigificantly faster, more accurate and works
across a range of fields-of-view (as opposed to just lumbar scans).
(2) Radiological grading adopts a more powerful architecture, adding several new grading 
schemes without loss in performance. A demo of the software is available at the project 
website: \texttt{http://zeus.robots.ox.ac.uk/spinenet2/}.
\end{abstract}

\begin{figure}[h!]
  \centering
  \includegraphics[width=0.9\linewidth]{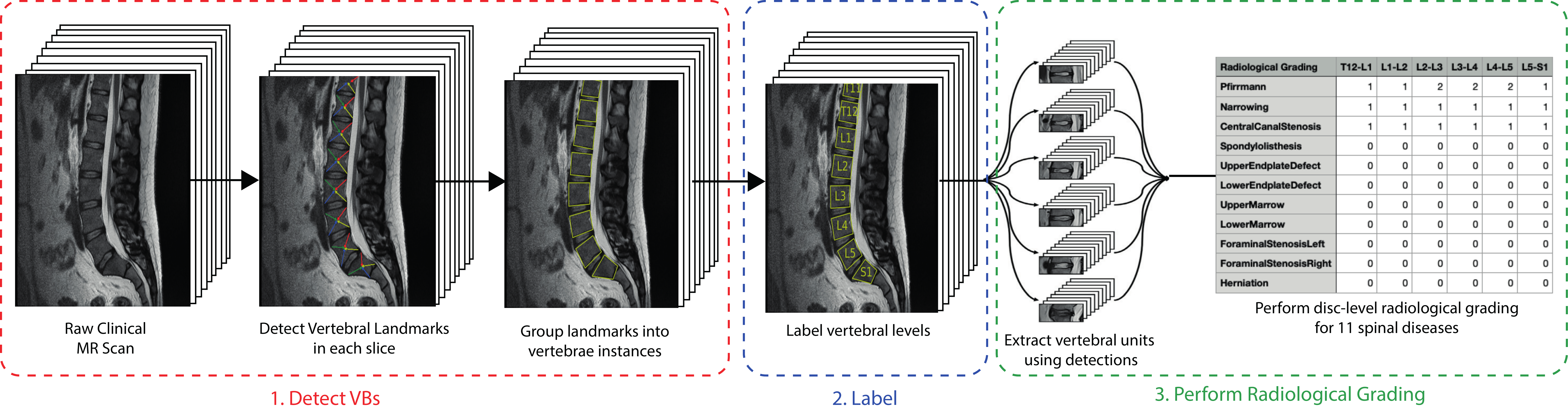}
  \caption{An overview of SpineNetV2, shown acting on a 
  T2 lumbar MR scan. Note that the vertebra detection and labelling pipeline
  works across a range of different MR sequences (T1, T2, STIR, etc.) and fields-of-view (e.g.\ cervical, thoracic, lumbar and whole spine) as illustrated in Figure~\ref{fig:example-detections}.}
  \label{fig:spinenet_overview}
\end{figure}

\newpage
\tableofcontents
\newpage

\section{Introduction}
\label{sec:introduction}

Back pain is the most common cause of long-term disability;
it affects around 80\% people in the UK during their lifetime~\cite{palmer_back_2000}. 
As people live longer, incidence will only increase. To combat this, we need methods to diagnose and monitor etiology 
like degenerative changes that are quick, effective and cheap to perform. 
This is the motivation behind SpineNet; to offer a completely automated set
of tools for performing common gradings \& measurements in clinical spinal 
MR scans.

This report describes the second iteration of the SpineNet software with
several improvements over the initial version\footnote{SpineNetV1 project page: 
\texttt{http://zeus.robots.ox.ac.uk/spinenet/}}. Namely, the method of detecting
and labelling vertebral bodies has been completely overhauled and is now much faster,
more robust, and can run across a range of 
fields of views (cervical, whole spine, lumbar, thoracic etc.)
as opposed to just lumbar scans. 
The grading network has also been improved to performs several new grading types; 
disc herniation, left and right 
foraminal stenosis and multiclass central canal stenosis (as opposed to 
binary classification in the original version). Grading performance across pre-existing
tasks is similar or slightly improved from SpineNetV1.

Crucially, unlike SpineNetV1 which was implemented using MATLAB, 
SpineNetV2 is implemented using open-source python libraries. Therefore, it can be run on a 
much wider range of hardware without the need for potentially expensive software
licences. The time-consuming, HOG-based vertebra detection system  of V1 has been replaced 
with a much faster (on GPUs and CPUs) and more robust deep-learning based approach.
Furthermore, the network used to perform radiological grading has now uses a more powerful
ResNet34 backbone to extract visual features as opposed to VGG-F. A complete summary of 
these changes can be seen in Table \ref{tab:changes}.

\begin{table}[ht]
    \begin{center}
        \renewcommand{\arraystretch}{2}
    \caption{Side-by-side comparison of the original SpineNetV1 and SpineNetV2.}
    \label{tab:changes}
    {\scriptsize
        \begin{tabularx}{\textwidth}{XX}\toprule
        \textbf{SpineNetV1} & \textbf{SpineNetV2} \\
        \toprule
        - Vertebra detection \& labelling on lumbar sagittal scans across a range of common clinical
        MR sequences&\
        - Vertebra detection \& labelling on any field of view sagittal scan\
        (e.g.\ cervical, lumbar, whole spine)\\
        - Implemented in MATLAB & - Implemented in open-source python libraries only (e.g. PyTorch, PyDicom)\\
        - Complete radiological grading of lumbar scan in approx. 5 minutes. &\
        - Complete radiological grading of lumbar scan in approx. 5 seconds.\\
        - Grades for (No.\ Classes): Pfirrmann (5), Disc Narrowing (4), Endplate Defects (2), Marrow Changes (2),\
        Spondylolisthesis (2) and Central Canal Stenosis (2).&
        - Grades for (No.\ Classes): Pfirrmann (5), Disc Narrowing (4), Endplate Defects (2), Marrow Changes (2),\
        Spondylolisthesis (2) and Central Canal Stenosis (\textcolor{red}{4}), \textcolor{red}{Foraminal Stenosis (2) and Disc Herniation (2)}.\\
        - VGG-F Backbone for Grading Network Backbone.& - ResNet34 Grading Network Backbone.\\
    \end{tabularx}
    }
    \end{center}
\end{table}

This technical report builds on several existing publications on the subject
of vertebrae detection and radiological grading, amongst which 
are~\cite{Windsor20b,Jamaludin17b,Lootus13,Lootus14,Jamaludin16,Jamaludin17}.

The report is organised as follows; 
Section \ref{sec:overview} describes the operation and functionality of SpineNetV2
at a high level. Section \ref{sec:methods} dicusses the methods
used for vertebral body (VB) detection (\S\ref{sec:vfr-detection}), labelling
(\S\ref{sec:convolutional-labelling}), extracting intervertebral volumes (IVVs)
from the detected VBs (\S\ref{sec:extract-ivd}), and performing radiological
grading of these detected IVVs (\S\ref{sec:grading}). 
Section~\ref{sec:implementation} discusses the implementation of the 
system, including the libraries (\S\ref{sec:code-implementation}) and 
datasets (\S\ref{sec:datasets}) used and methods
for training the consituent neural networks (\S\ref{sec:training}).
Section~\ref{sec:results} gives experimental results for the detection and labelling
pipeline as well as the radiological grading pipelines, including comparisons with the 
original version of SpineNet. Finally, Section~\ref{sec:conclusions} concludes
the report, giving future plans for SpineNet as well as information on how SpineNet
software can be used.

\section{Overview}
\label{sec:overview}

SpineNetV2 provides an entire pipeline to go from raw DICOM files 
(as output by a conventional MRI machine) to 3-D detections
localising the each vertebral body visible in the scan and labels
describing the level of each detection. In the case
of T2 lumbar scans, the pipeline can also output multiple common radiological
gradings for each intervertebral disc visible in the scan.

\paragraph{Inputs:} DICOM files corresponding to slices of
a sagittal MRI spinal scan. This scan can be one of a range of common clinical 
sequences (e.g.\ T1, T2, STIR, FLAIR, TIRM, Dixon-technique, etc.), 
have an arbitrary field-of-view (e.g.\ lumbar, cervical, whole spine, etc.)
and be of varying resolution and slice thickness.

\paragraph{Outputs:}
\begin{itemize}
    \item CSV/JSON file describing the location of vertebral bodies in the scan and the 
    corresponding level of these detections (from S1 to C2).
    \item \textit{T2 Lumbar Scans Only}: CSV/JSON file 
    with predictions for a range of common radiological grading schemes 
    (itemized
    in Section~\ref{sec:grading}) for each intervertebral disk visible in the
    scan. 
\end{itemize}

\subsection{Processing Pipeline}
SpineNet consists of multiple processing steps, each of which is described
in detail in the following sections. These are:

\begin{enumerate}
    \item Splitting each slice of the input DICOM file into a set of smaller 
    patches (for large scans only).
    \item Detecting each visible vertebral landmark (vertebral corners \& centroid) 
    in each patch/slice.
    \item Grouping together these landmarks into vertebrae instances.
    \item Determining the level of each detected vertebra.
\end{enumerate}

\begin{figure}
    \centering
    \begin{subfigure}{\textwidth}
        \centering
        \includegraphics[width=.9\textwidth]{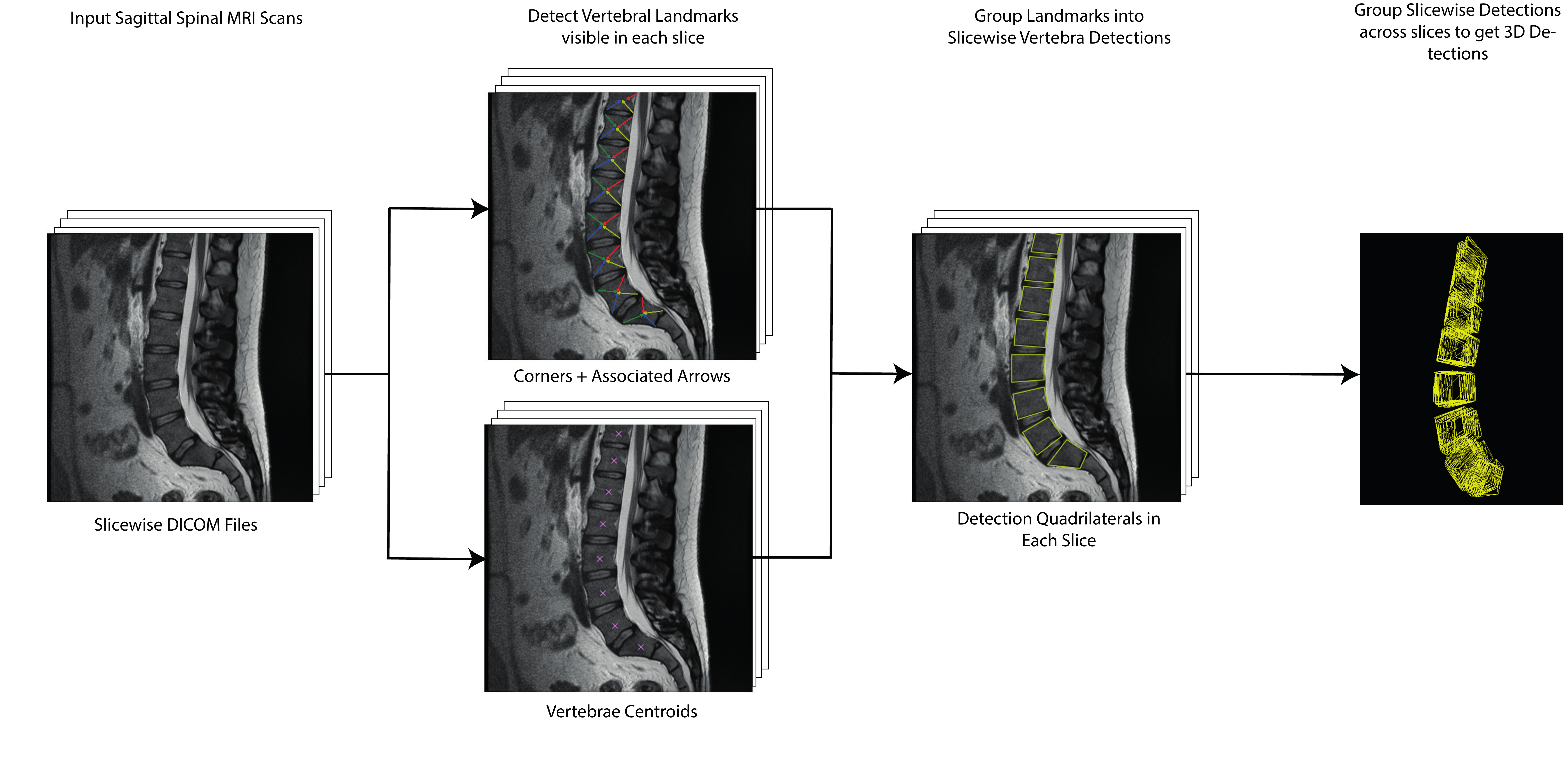}
        \subcaption{\textit{VB Detection} - Detecting 3D bounding boxes around each visible vertebral body.
        Larger scans are split into multiple patches, as explained
        in Section~\ref{sec:patch-splitting}.}
    \end{subfigure}
    \begin{subfigure}[b]{\textwidth}
        \centering
        \includegraphics[width=.9\textwidth]{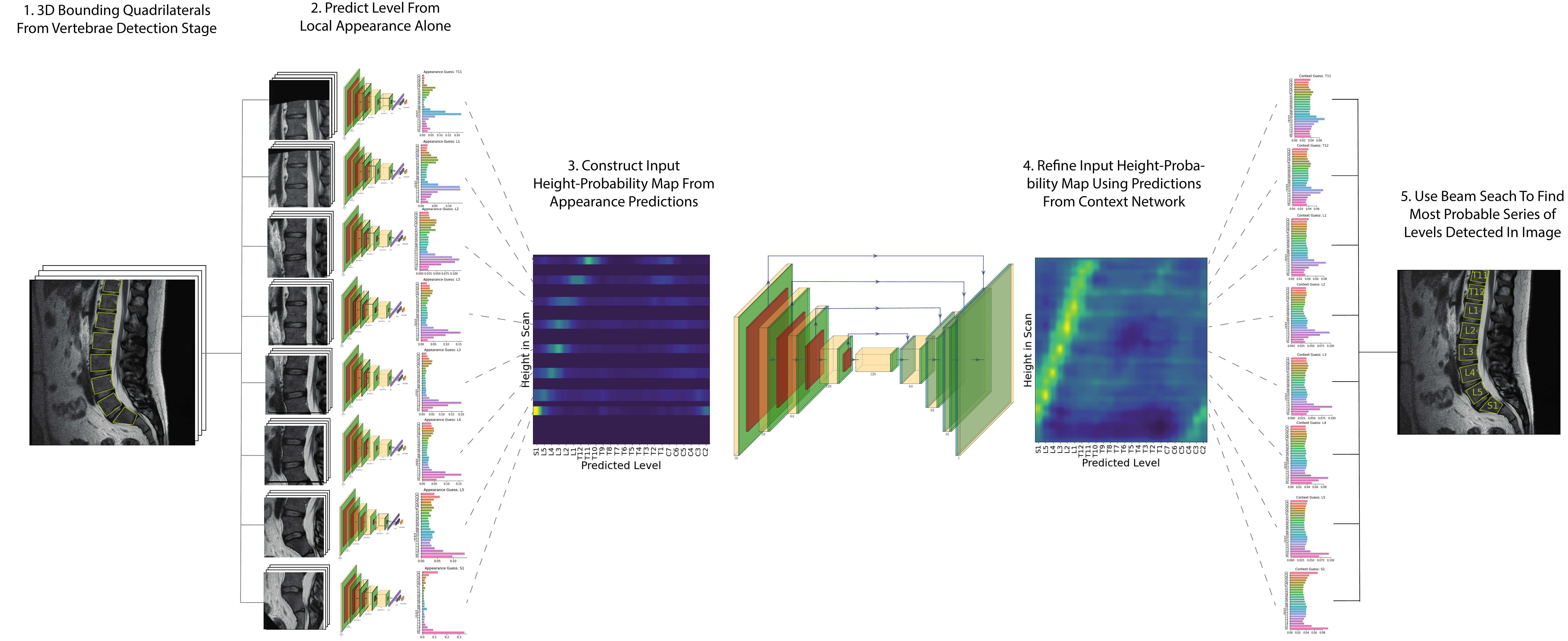}
        \subcaption{\textit{VB Labelling} - Determining the level of each detected vertebra.}
    \end{subfigure}
    \begin{subfigure}{\textwidth}
        \centering
        \includegraphics[width=.9\textwidth]{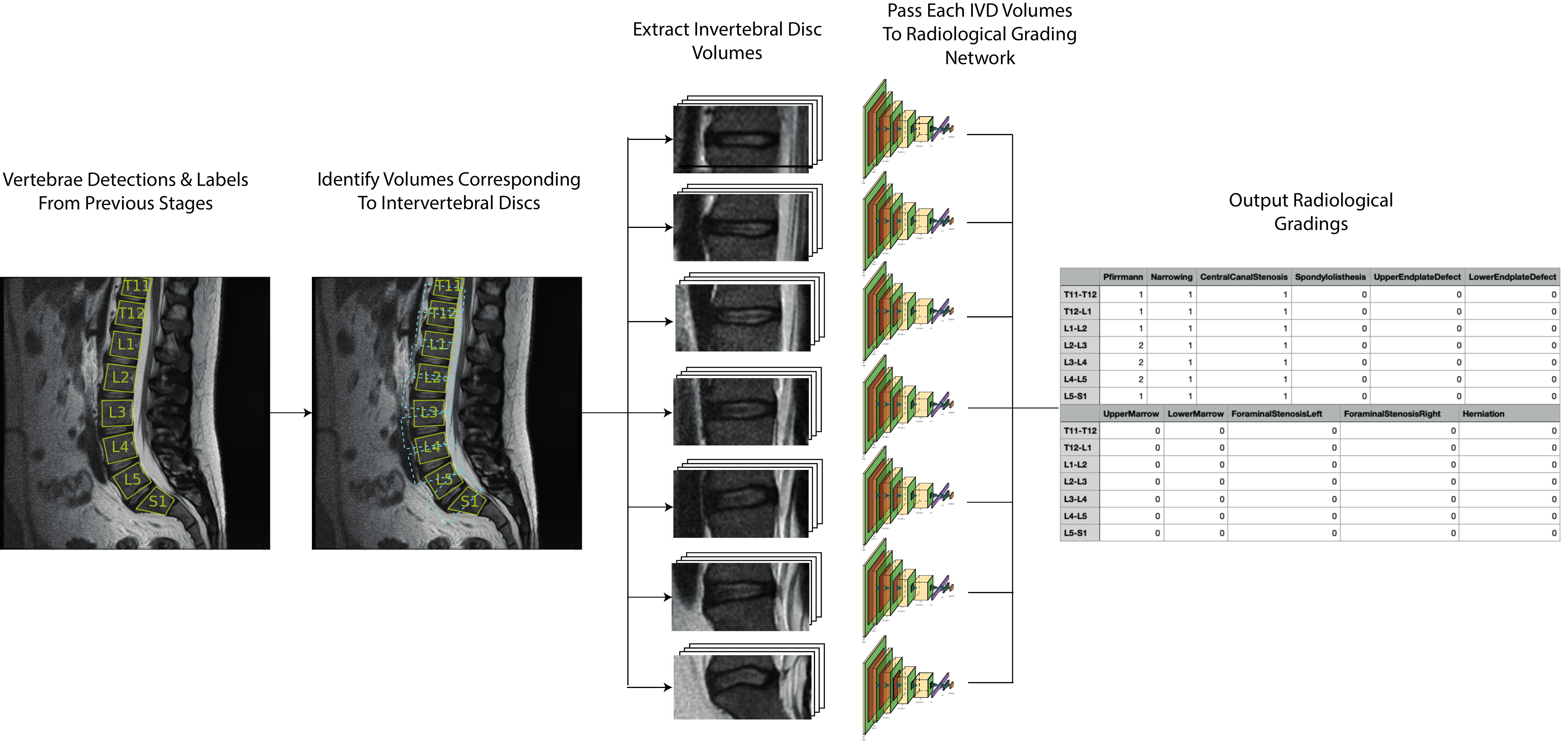}
        \subcaption{\textit{VB Grading} - Grading each detected IVV  using
        a range of common radiological
        grading schemes. Note this step only applies to T2 Lumbar scans.}
    \end{subfigure}
    \caption{The processing pipeline for SpineNetV2. Note that the vertebra 
    detection and labelling section works across a range of sequences 
    and fields-of-view, whereas the radiological grading system only 
    operates on T2 lumbar scans.}
    \label{fig:processing-pipeline}
\end{figure}

For T2 Lumbar scans the following stages are added:

\begin{enumerate}
    \setcounter{enumi}{4}
    \item Locating IVVs (inter-vertebral volumes) using the VB detections.
    \item Performing radiological grading of the extracted IVV.
\end{enumerate}
This entire process is shown in Figure~\ref{fig:processing-pipeline}. 

\newpage

\section{Methods Used}
\label{sec:methods}

This section decribes methods used for the three main stages of the SpineNet
pipeline (VB detection, VB labelling and IVV grading) from
an algorithmic standpoint. 
These stages are illustrated in Figure~\ref{fig:spinenet_overview}. 
Broadly speaking, whole spine scans and smaller scans (e.g.\ lumbar, cervical)
are processed identically, with only minor differences in how the patch splitting 
is done in the VB detection stage. These differences are described fully in Section~\ref{sec:patch-splitting},
however we will initially assume a single lumbar scan when describing the processing pipeline.

\subsection{VB Detection by Vector Field Regression}
\label{sec:vfr-detection}

The first stage in the SpineNetV2 pipeline is to detect vertebral bodies (VBs) in the
raw scans. This is done using a method called \textit{Vector Field Regression} (VFR). Operating
in each sagittal slice of the scan, a network detects gaussian responses over 
the centroids and corners (\textit{vertebral landmarks}) of each visible 
VB. For each corner detected, a corresponding vector field is also
output. At the point of each corner, this corresponding vector field should `point' to the 
centroid of the vertebra to which it belongs. This allows landmarks from the
same vertebra to be grouped together in a simple manner which is robust to rotations,
flips and variable vertebra size. 

\begin{figure}[h] 
  \includegraphics[width=\linewidth]{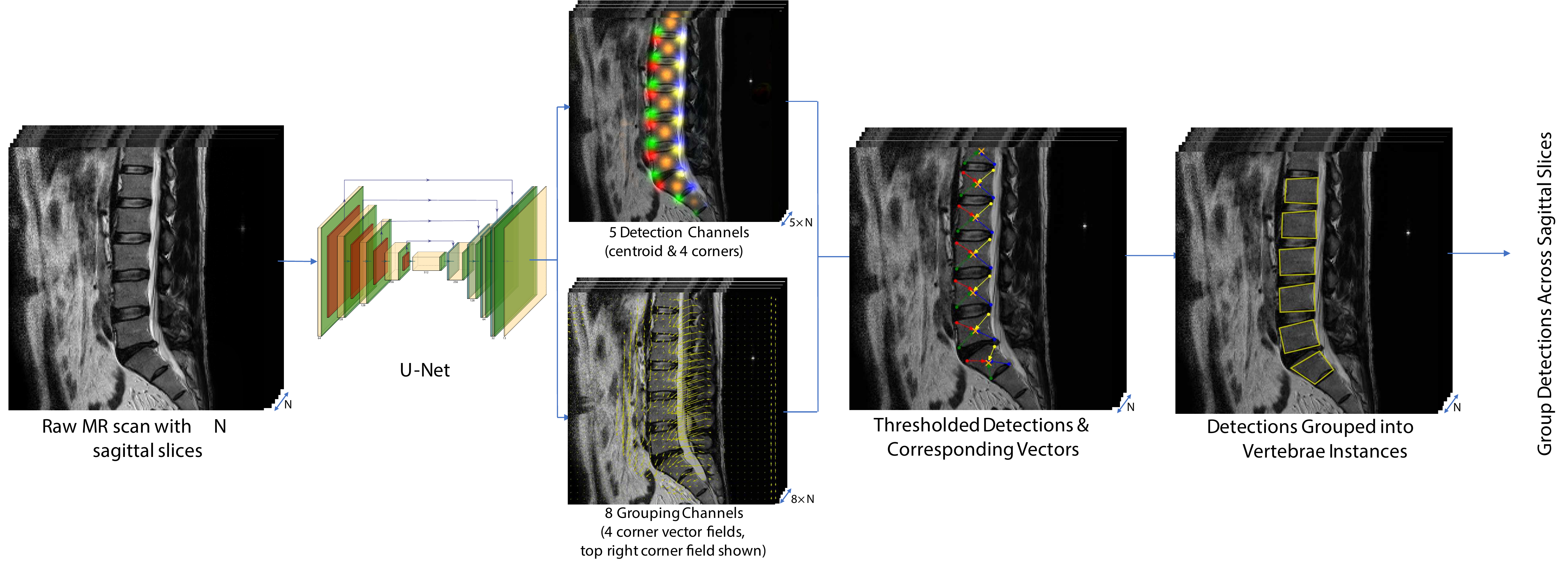}
  \caption{An overview of the VFR (\textit{Vector Field Regression}) method. A U-Net has 13 output channels;
  5 detection channels which output gaussian heatmaps over the four corners
  and centroid of each visible vertebral body and 8 grouping channels; the 
  $x$ and $y$ components of vector fields corresponding to each corner.}
  \label{fig:vfr}
\end{figure}

At this point, each detection consists of a centroid and four corners grouped together 
to make a quadrilateral in an individual sagittal slice. Slicewise 2D quadrilaterals corresponding
to the same vertebra must then be grouped together across slices forming 3D volumes.
This is done are by measuring the intersection-over-union (IOU)
of quadrilaterals in neighbouring slices. If the IOU is large, the quadrilaterals are
grouped together into a single vertebral instance.
A diagram of this process going from raw scans to 3D volumes is shown in Figure~\ref{fig:vfr}.

Specifically, inference proceeds as follows:
\begin{enumerate}
  \item A raw MR scan of dimension $S \PLH H \PLH W$ is split into $S$ sagittal slices.  These slices are in turn split into 
  square patches of size 25 cm\textsuperscript{2} with an overlap of 30\% between 
  neighbouring patches on each side. This patch-splitting allows the network to
  deal with scans of varying dimensions.
  \item The patches are resampled by cubic interpolation 
  to resolution of $224\PLH 224$. The resamples patches are 
  fed as input to a ResNet50-encoded U-Net. This results in a 
  $13\PLH 224\PLH 224$ output per patch; 5 detection channels 
  and 8 grouping channels (see Figure~\ref{fig:vfr}).
  \item By resampling \& concatenating these patch-level outputs using the median output
  for overlapping regions, an slice-level output tensor is reconstructed of size
  $13\PLH H\PLH W$.
  \item The detection channels are then decoded to find the location of detected
  VB centroids and corners. This is done by thresholding each channel \& finding all 
  connected component in the resulting binary map. The exact point-of-detection
  for each landmark is the point of maximal response in the detection channel
  for the corresponding connected component.
  \item Now that each VB landmark in the slice has been detected, they must be 
  grouped into quadrilaterals corresponding to individual VBs. This is done by
  measuring the value of the corresponding vector field at the point-of-detection
  for each corner landmark. Looping through each detected VB centroid, the corner
  landmark of each type which points closest is assigned as the corresponding
  corner for that centroid, forming a quadrilateral. If no arrow is 
  within a distance from the centroid
  of 50\% of the arrow length, that centroid is discarded as a spurious detection.
  \item Finally the resulting VB polygons are grouped across neighbouring slices if they have an
  IOU over 0.25. If more than one polygon in a neighbouring slice overlap, 
  then the one with the greatest IOU is chosen.
\end{enumerate}

\subsubsection{Splitting Large Scans Into Patches}
\label{sec:patch-splitting}
\begin{figure}
  \includegraphics[width=\linewidth]{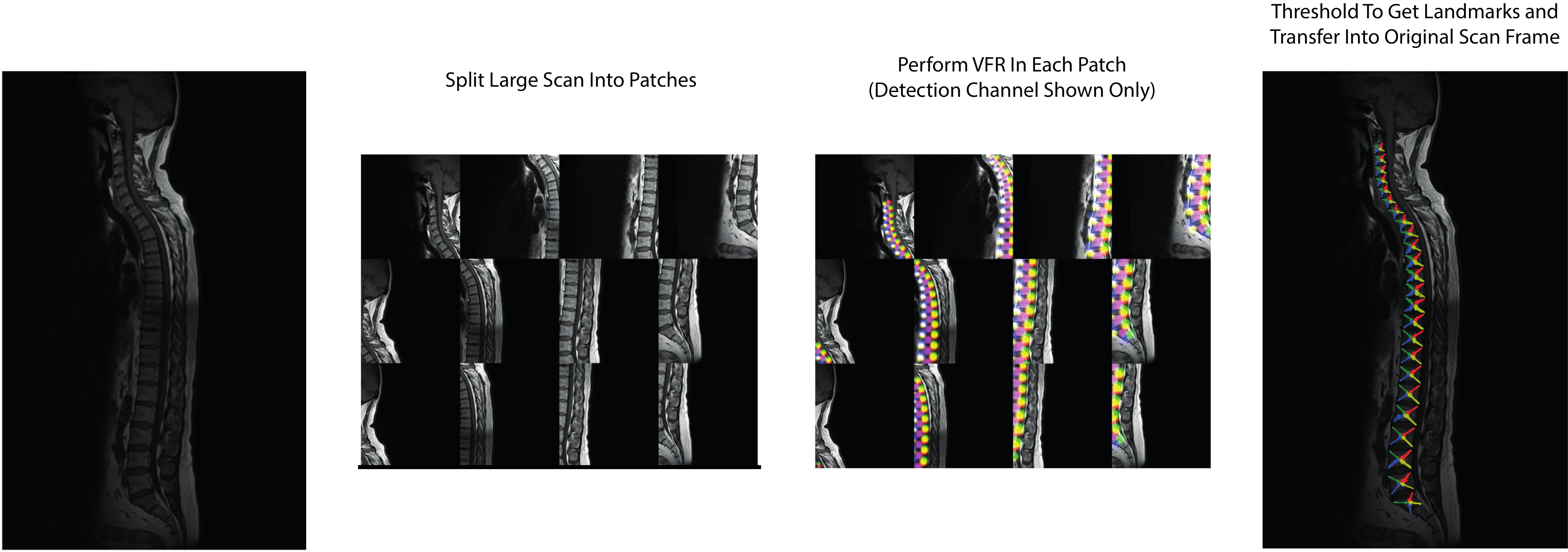}
  \caption{Splitting a large scan into patches before applying VFR.}
  \label{fig:patch-splitting}
\end{figure}

Larger non-square scans (such as whole spine scans) are split into patches 
before VFR is applied. This is done by splitting the scan into a grid of
overlapping squares with edge length $50$cm (as determined by the pixel
spacing parameter in the DICOM header) and an overlap of 40\% between 
neighbouring patches. The output from the detection and grouping channels
are then used to find landmarks in each patch.
These landmarks are then transformed back into the frame of the original scan. 
At this point, the algorithm proceeds as before by grrouping landmarks into slicewise
polygons and then across saggital slices.
Figure~\ref{fig:patch-splitting} shows the process of patch-splitting
for a whole spine scan. 

\subsection{Convolutional Labelling of Vertebral Levels}
\label{sec:convolutional-labelling}
The VFR method described above allows us to detect vertebral bodies in a sagittally 
sliced MR scan. The next step of the pipeline is to determine which vertebral
levels the detections correspond to (e.g.\ S1, L5, L4 etc). 
This is made more challenging by the fact that SpineNetV2 is not constrained
to a single field of view. As such, `counting up' methods, i.e. those that rely 
on an anchor vertebra being visible (such S1/C2 at the bottom/top of the vertebral
column) are unsuitable. Furthermore, such methods are not robust to missed detections
or variations in the number of vertebra in the vertebral column (e.g.\ in cases
where a transitional vertebra is present).

There are two pieces of information to consider when labelling a vertebra - its
\textit{appearance} (e.g.\ intensity pattern, shape, size etc.) and its 
\textit{context} (the vertebra's position relative to other detections in the 
scan). For example, S1 usually has a very distinctive shape which allows it to
be labelled from appearance alone. On the other hand, L5 looks very similar 
to the other lumbar vertebrae, however can be easily identified from it's 
context - it is the next vertebra up from S1. Our method attempts to use both
these sources of information when labelling a vertebra.

\begin{figure}[h]
  \includegraphics[width=\linewidth]{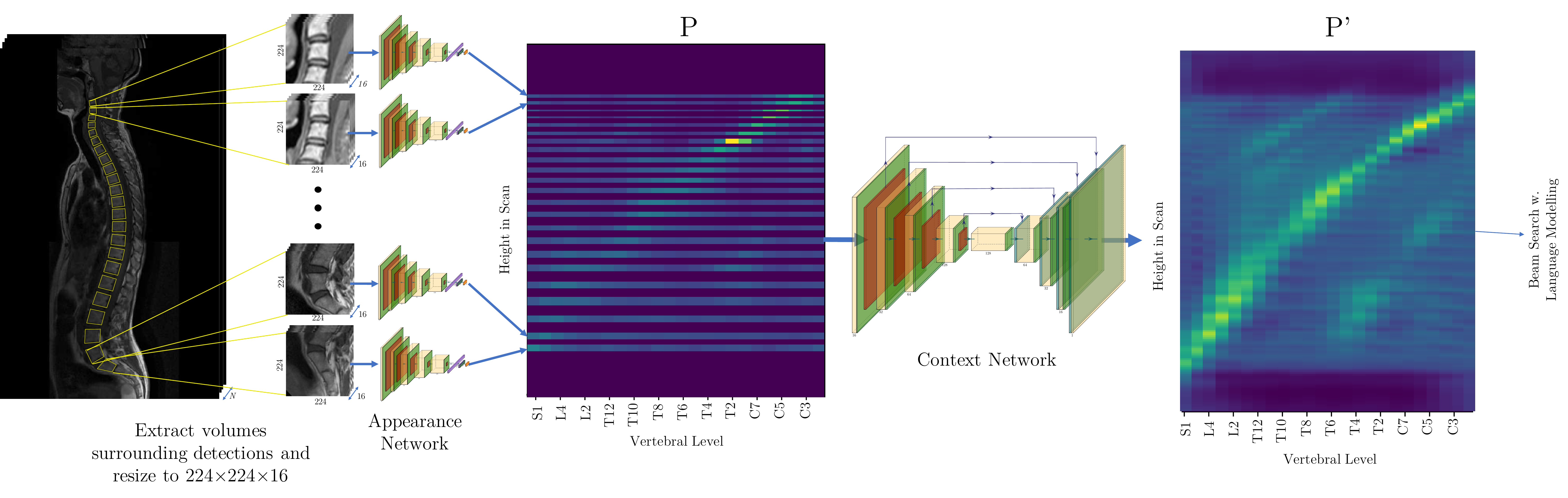}
  \caption{The language modelling-inspired approach to vertebra labelling
  outlined in this section. We begin by constructing a probability-height map
  $P$ based on appearance information alone. Using a context-aware CNN this map
  is refined ($P'$). Finally $P'$ is decoded into a valid sequence of levels 
  using a penalised beam search.} 
  \label{fig:labelling-pipeline}
\end{figure}

Firstly, a 3D volume around each detected VB is fed as input to an \textit{appearance
network}. This outputs a 23-element (from S1 to C3) probability vector predicting the level 
of the VB from appearance alone. To include information about the spatial
configuration of the detections, this is then used to contruct a probability-height 
map, $P$. At the height of each detection, $P$ has value equal to the output from
the appearance network for that detection. Using this as input, a convolutional \textit{context
network} refines the height-probability map, taking into account appearance
predictions from spatially nearby VBs to update the probability vectors for each 
detection. The result is a refined height-probability map $P'$. 
This process is shown in Figure~\ref{fig:labelling-pipeline}. Finally, $P'$ needs 
to be decoded into discrete level predictions. Na\"ively, this could be done by   
taking the maximum probability level at the height of each detection in $P'$.
However, this would allow for nonsensical outputs, such as the same level for
two detections. Ideally, we also want to build in soft contraints such that
successive detections are labelled as successive labels. For example, we would
expect S1 to be the detection below L5. However, we also wish to remain robust to 
missed detection.

To build these constraints into our approach, we take inspiration 
from language modelling. Using a beam search, we can find the most probable
valid sequence of levels for the detections. Penalties are added to the sequences 
probability score in the case of transitional vertebrae or numerical variations to
reflect the unlikeliness, yet possibility, of such events. The exact parameterisation
of this search is specified in the next paragraph.

The exact specification of this labelling stage is as follows:
\begin{enumerate}
  \item Given VB detections from the previous stage, a 2-D bounding 
  cube is fit tightly around the union each of detection's slicewise 
  series of polygons. By expanding this cube by 100\% on each side in the
  axial and coronal directions and by 50\% on each side in the sagittal 
  direction a volume around each detection and its nearby anatomical features
  is created. 
  \item All extracted volume are resized by cubic interpolation to $224\times224\times16$ 
  (where 16 is the number of sagittal slices). These are then fed to an 
  appearance network which outputs a 23-element probability vector, attempting
  to classify the VB as a level from S1-C3.
  \item The output probability vectors are 
  re-calibrated~\cite{guo_calibration_2017} using a softmax temperature $T=10$. 
  These recalibrated vectors are then used to construct an 
  initial probability-height map,
  $P\in \mathbb{R}^{H\times23}$. For a detection polygon with output 
  appearance probability vector $\mathbf{p}_a\in \mathbb{R}^{23}$ and 
  spanning from height $h_1$ to $h_2$ in the scan, $P$ is equal to this value 
  between these heights, i.e.
  $P(h)=\mathbf{p}_a \:\forall\: h_1 \leq h \leq h_2$. 
  \item $P$ is then given as input to a image-to-image \textit{context} CNN.
  This outputs $P'\in\mathbb{R}^{H\times23}$, a refined probability-height
  map which considers the neighbours of each detection to update its probability
  vector. This refinement step can be seen in Figure~\ref{fig:labelling-pipeline}. 
  \item The next step is to decode $P'$ into a series of discrete level predictions.
  Firstly, the probability vector at the centroid of each detection, $\mathbf{p}_i$
  is extracted. The joint probability of a sequence of levels can then be calculated 
  as the product of elements of these vectors. For example, if detected 
  VB $i$ has a probability of being S1 of $a$ and VB $j$ has probability
  of being L5 of $b$, then the joint probability of the level sequence
  is $a\times b$. To reduce search space, we use a beam search to find the most 
  likely sequence, starting from the bottom detection and only storing the 100
  most probable level sequences at each step.
  \item To impose constraints on sequences of level predictions, penalties can 
  be added to the sequence scores. For example, sequences which predict the same
  level twice are given a probability of 0. Furthermore, if there is a missed 
  detection (i.e. L4 directly above S1), the joint probability score is 
  multiplied by a penalty score to reflect the rarity of this event. 
  To allow for cases where S2 or C2 are detected, double detections of S1 or C3 
  are allowed without a score penalty and then relabelled in post-processing. 
\end{enumerate}

At this point, each vertebra in the scan from S1 to C3 should have been detected
and assigned a level label. These detections can be used in a range of applications.
For example in whole spine scans, SpineNet can be used to measure spinal curvature
(for scoliosis measurement) or extract regions of anatomical interest such as the
spinal chord or vertebral bodies for lesion segmentation (e.g.\ spinal metastases 
or ankylosing spondylitis). 

One area of particular interest is radiological grading
of degenerative changes. The following section describe how this is done in
the context of the SpineNet.

\subsection{Extracting Intervertebral Volumes}
\label{sec:extract-ivd}
To perform radiological grading in T2 lumbar scans, first volumes surrounding each 
intervertebral disc must be extracted. 
This is done using the VB detections from the previous stage.
The mid-point between the centroids of two consecutive VB detections is calculated.
This defines the centre of each extracted IVV. From this the volume
is rotated such that the lower endplate of the upper vertebra is horizontal. The
width of the extracted endplate volume is then defined as double that of the
larger VB detection. The height of the IVV is then chosen such that the aspect
ratio of the extracted patch is 2:1. This is done across all slices in which
the vertebrae is detected.

\subsection{Radiological Grading of Intervertebral Volumes}

Once the IVVs are extracted, they are resampled to have a resolution of
$112\times224\times9$ (height, width and sagittal slices respectively) 
These volumes are then fed to a radiological grading network which outputs 
scores for the following
radiological gradings:
\begin{enumerate}
  \item Pfirrman Grading (5 classes)
  \item Disc Narrowing (4 classes)
  \item Central Canal Stenosis (4 classes)
  \item Upper \& Lower Endplate Defects (Binary)
  \item Upper \& Lower Marrow Changes (Binary)
  \item Left \& Right Foraminal Stenosis (Binary)
  \item Spondylolisthesis (Binary)
  \item Disc Herniation (Binary)
\end{enumerate}

The radiological grading network is a standard multi-class classification 3D CNN.
We experiment with a range of different architectures, described in Section~\ref{sec:results}.
Each of these grading schemes is described in further detail in Section \ref{sec:datasetsGenodisc}.
Note that the same model is used for all gradings schemes and vertebral levels.

\label{sec:grading}

\section{Implementation}
\label{sec:implementation}

This section describes the specific design choices made while developing SpineNetV2,
as well as details of the datasets and methods used to train the constituent
neural networks.

\subsection{Code Implementation}
\label{sec:code-implementation}
All stages of the SpineNetV2 processing pipeline are implemented entirely using open-source
python libraries. Deep learning functionality is provided by 
CUDA-enabled PyTorch v1.7 (or later). Input DICOMs are processed using pydicom.

The software is designed such that detection and grading can be seperated. During
inference, models can run on CPUs or GPUs for faster processing (the performance and
memory constraints of each configuration are given in Section~\ref{sec:processing-speed}).

\subsection{Model Architectures}
There are 4 constituent neural networks in the SpineNetV2 pipeline; the VFR regression network,
the appearance network, the context network and the grading network. The VFR regression
network is a ResNet18-encoded UNet. The appearance network is a simple VGG-F network as outlined
in the original paper~\cite{chatfield_return_2014}. The context network is a simple 
UNet~\cite{ronneberger_u-net_2015}. Finally, the radiological grading network is
a conventional ResNet32 model with 3D $3\PLH3\PLH3$ convolutions in the first layer
and $3\PLH3$ convolutions in all other layers.

\subsection{Datasets}
\label{sec:datasets}
Two datasets are used to train SpineNetV2; \textit{Oxford Whole Spine}(OWS) and 
\textit{Genodisc.} OWS is a 
dataset of whole spine scans across a range of commonly
used sequences, used only for training the vertebra detection and labelling sections of the 
SpineNetV2 pipeline.
\textit{Genodisc} is a dataset of sagittally-sliced lumbar T2 scans used 
in training all stages of the pipeline.

\subsubsection{OWS}
\label{sec:datasetsOWS}
\begin{figure}
  \begin{subfigure}{0.49\textwidth}
    \centering
    \includegraphics[width=\textwidth]{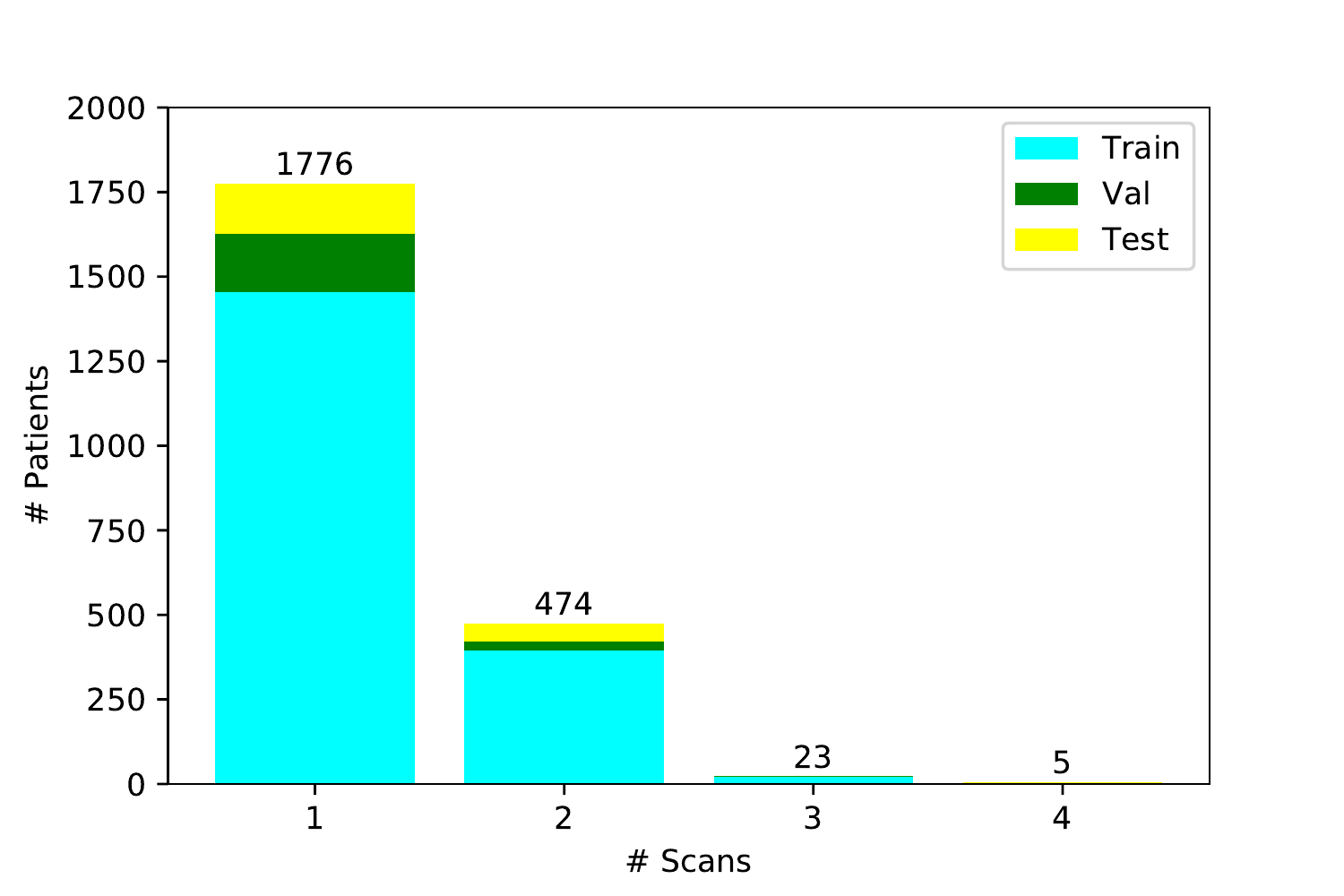}
    \caption{Scans per patient in Genodisc.}
    \label{fig:genodisc-num-scans-dist}
  \end{subfigure}
  \begin{subfigure}{0.49\textwidth}
    \centering
    \includegraphics[width=\textwidth]{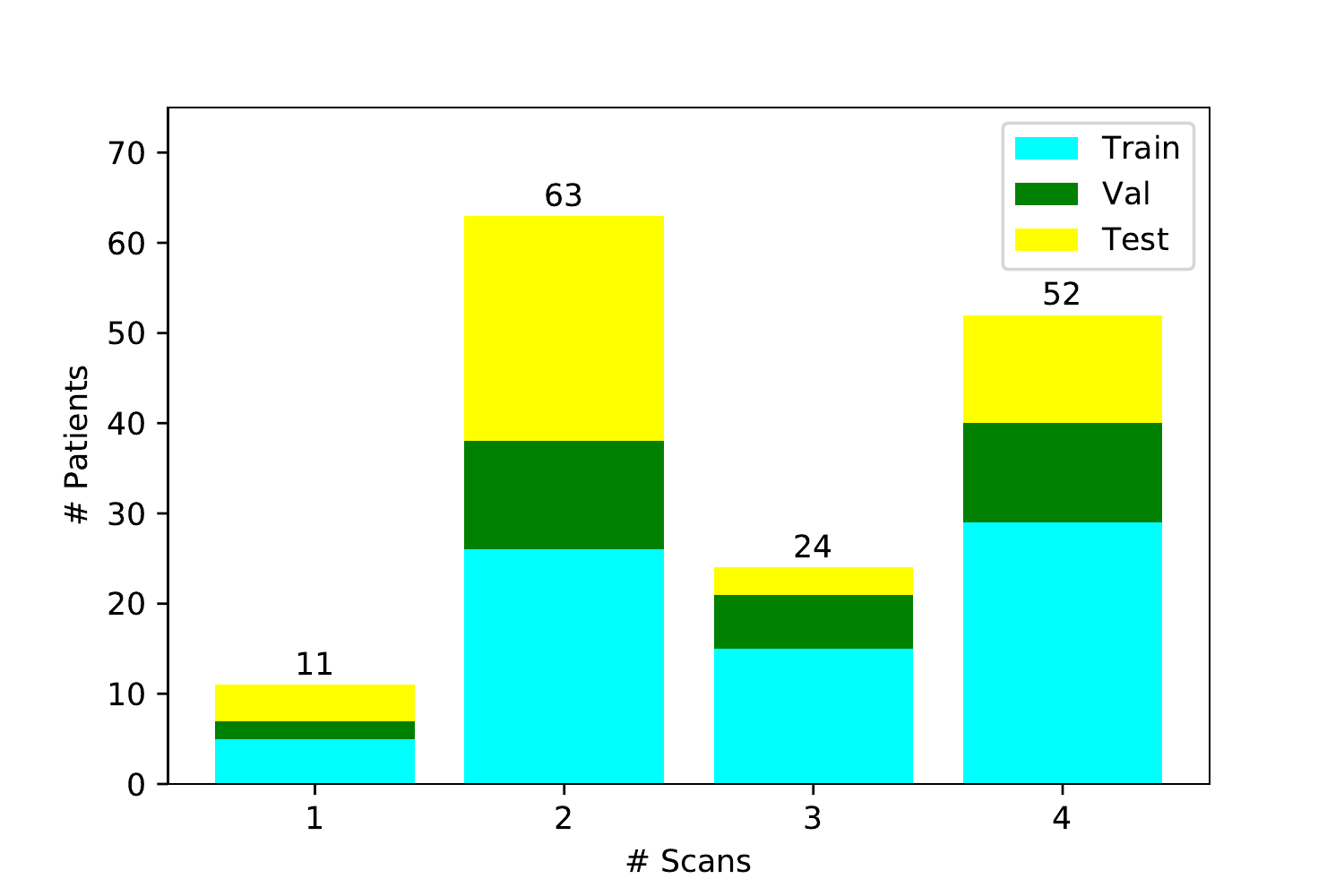}
    \caption{Scans per patient in OWS.}
    \label{fig:ows-num-scans-dist}
  \end{subfigure}
  \begin{subfigure}{\textwidth}
    \centering
    \includegraphics[width=0.49\textwidth]{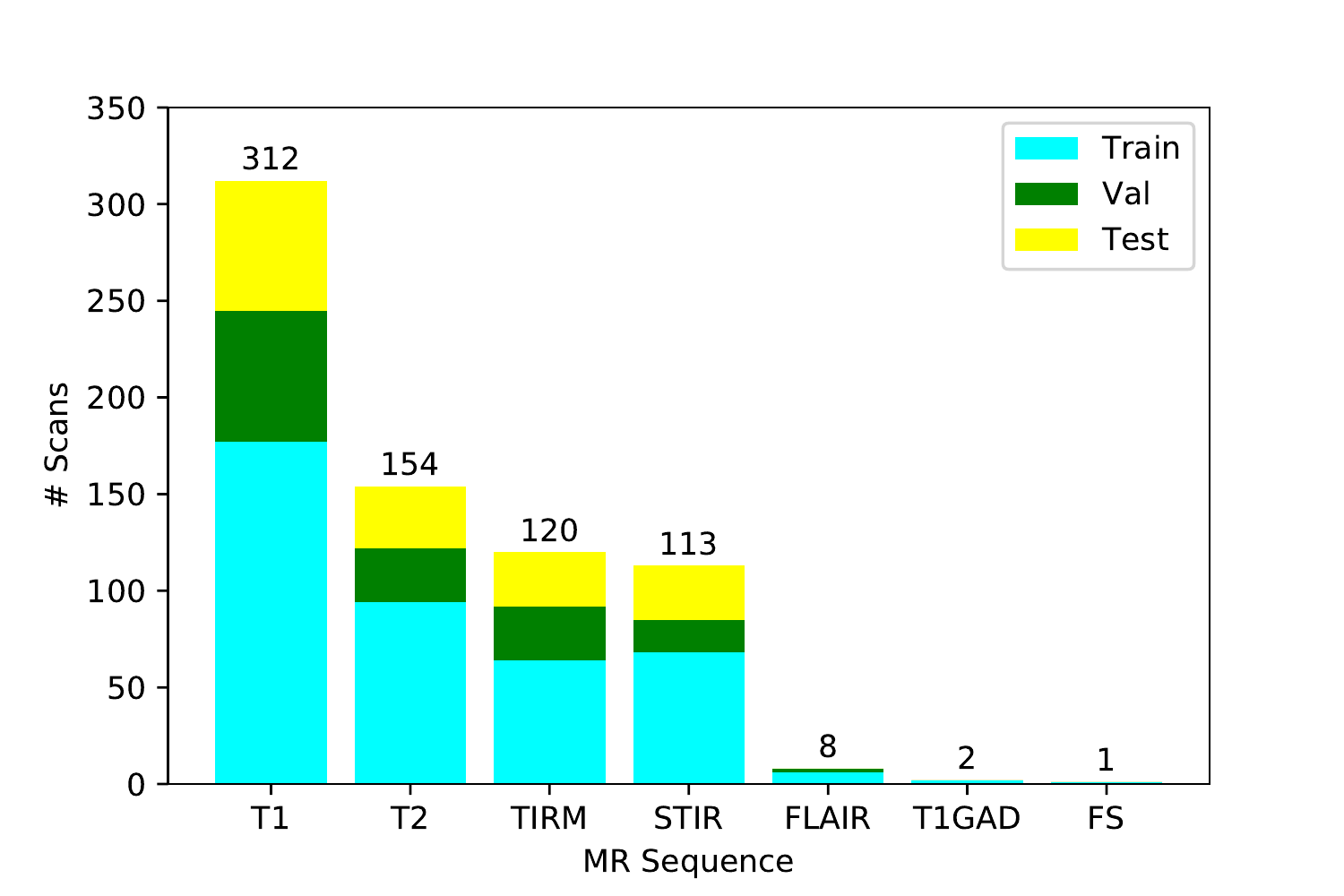}
    \caption{MR sequence types in OWS.}
    \label{fig:ows-seq-dist}
  \end{subfigure}
  \caption{Breakdown of scan types in both OWS (196 patients, 719 scans) and 
           Genodisc (2279 patients, 2819 scans). Both datasets are split 80:20:20\% 
           down the patient line.}
            
\end{figure}

OWS consists of 710 sagittally-sliced whole spine scans across 196 patients
extracted from local orthopaedic centre's PACS (picture archiving and communication system) and anonymised under
appropriate ethical clearance. These scans are across a range of commonly used clinical sequences (mostly T1, T2, STIR and TIRM). The distribution of sequences can be seen
in Figure~\ref{fig:ows-seq-dist}. Each vertebral body from S1 to C2 is annotated 
as a quadrilateral in the central slice of each scan by a non-specialist. 
These are used to generate the ground truths for training, as discussed in 
section~\ref{sec:training}.

\subsubsection{Genodisc}
\label{sec:datasetsGenodisc}
Genodisc is a dataset of sagittaly-sliced lumbar T1 and T2 scans
from 6 different international clinical spinal imaging centres. This dataset
is used for training the detection (T1 \& T2) and the radiological grading (T2 only)
stages of the pipeline. Each T2 scan is annotated by an expert radiologist. The
following degenerative changes are graded:
\begin{enumerate}
  \item Pfirrman Grading (5 classes) - A general grading system proposed in 2007 
  by Griffith et al. to categorise the degree of intervertebral disc (IVD) degeneration (originally for 
  older patients). Ranges from 1 (no degeneration) to 5 (severe degeneration)%
  \item Disc Narrowing (4 classes) - The width of the IVDs. Ranges from 1 
  (no narrowing) to 4 (extreme narrowing).
  \item Central Canal Stenosis (4 classes) - A narrowing of the spinal canal which
  can in turn lead to compression of the spinal chord. Ranges from 1 (no compression),
  2 (mild compression), 3 (moderate compression) and 4 (severe compression).
  \item Upper \& Lower Endplate Defects (Binary) - Abnormalities/damage to the
  top or base of the VB's constituent bodies. 
  \item Upper \& Lower Marrow Changes (Binary) - Lesions/changes in the intensity of
  constituent VBs.  
  \item Left \& Right Foraminal Stenosis (Binary) - A narrowing of the intervertebral
  foramina (openings where spinal nerves leave the central canal). In severe cases,
  this can lead to nerve compression similar to central canal stensosis.
  \item Spondylolisthesis (Binary) - A condition when a vertebra slips forward onto
  the vertebral disc below. This is often caused by a fracture in the 
  \textit{pars interarticularis}, a segment of bone that joins the vertebrae.
  \item Disc Herniation (Binary) - A condition where the centre of the IVD 
  (\textit{nucleus pulposus}) breaks through its casing (\textit{annulus fibrosus}).
  This can lead to nerve compression.
\end{enumerate}

\begin{figure}
  \centering
  \includegraphics[width=0.8\linewidth]{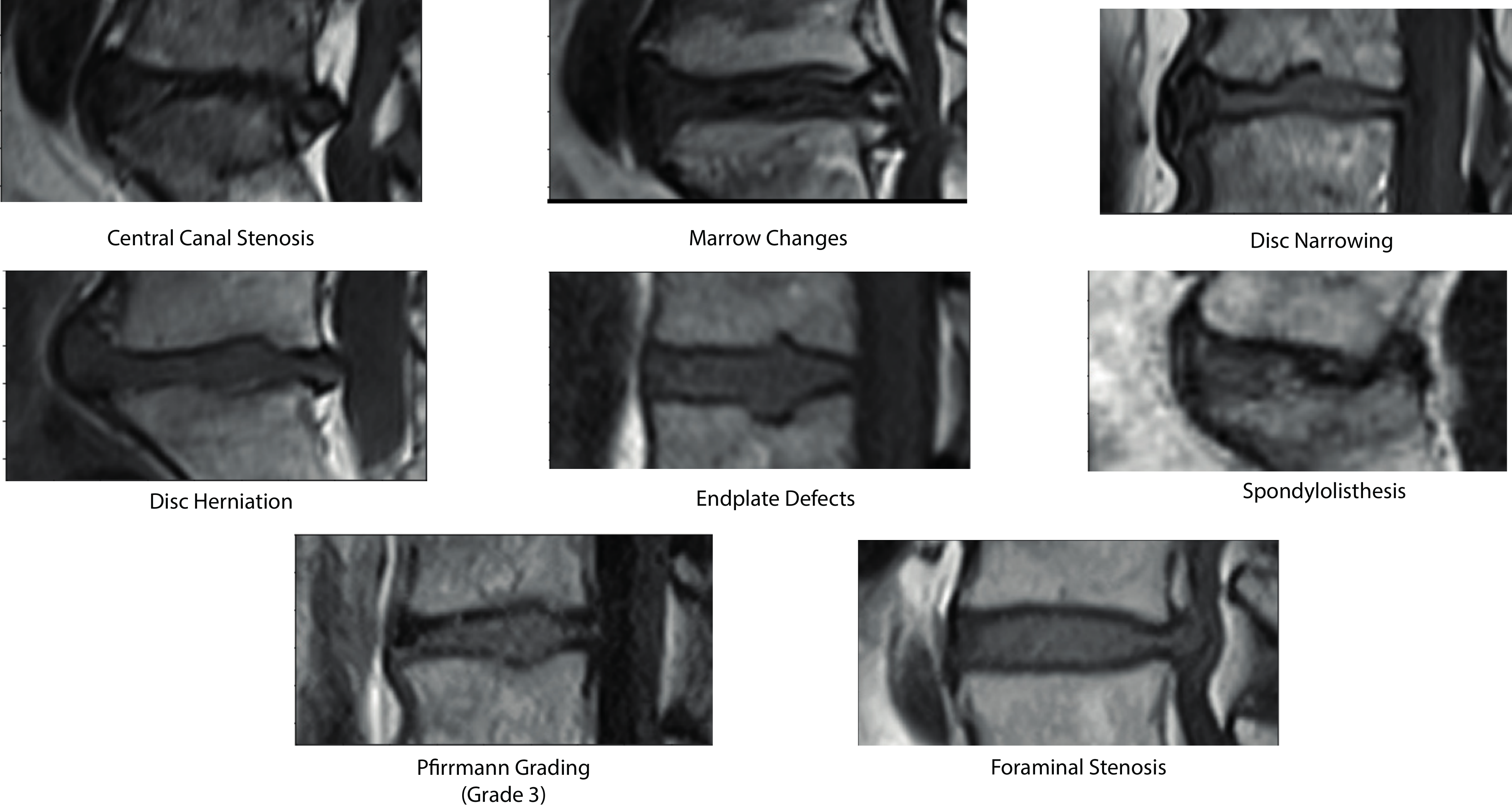}
  \caption{Example degenerative changes from the Genodisc training dataset.}
  \label{fig:example-degenerative-changes}
\end{figure}

Examples of these degenerative changes from the training dataset are shown in 
Figure~\ref{fig:example-degenerative-changes}.

\subsection{Training}
\label{sec:training}

In total 4 networks are trained for the SpineNetV2 pipeline: 1) a detection
network which detects vertebral bodies agnostic to their level; 2) a labelling 
appearance network which aims to label vertebrae based on their appearance alone;
3) a labelling context network which refines the predictions of the appearance 
network based on the appearance of neighbouring VB detections 4) a radiological 
grading network which operates on volumes surrounding IVDs and outputs gradings
for the degenerative changes listed in section~\ref{sec:datasetsGenodisc}.
Each network is trained independently on data from either Genodisc, OWS or
a combination of both.

\subsubsection{VB Detection}

The network is trained on patches of sagittal slices from Genodisc and OWS.
These patches cover an area of approximately 25cm\textsuperscript{2} and are 
resampled to $224\PLH 224$ pixels via cubic interpolation.  

Ground truths are constructed from VB polygons marked by annotators. The
network's 13 output channels are divided in two types: \textit{detection channels}
 which indicate the location of individual landmarks in the image and \textit{grouping channels}
 that show which vertebra each landmark belongs to by `pointing' to the corresponding centroid. The
 ground truth for each of these two channel types is generated as follows:
\begin{itemize}
  \item \textit{Detection Channels:} A gaussian response is added to the 
corresponding channel for each VB vertex, normalised to have 
a peak value of 1 and have variance proportional to the VB polygon's surface area.
A gaussian response is added to the centroid detection channel at the centre of
each annotated polygon.
  \item \textit{Grouping Channels:} The grouping channels for each vertex
are constructed such that, for an area around each vertex proportional to 
the VB's surface area, the two corresponding grouping channels together point 
to the VB's centroid.
\end{itemize}

Once the target tensor, $\hat{Y}$, is constructed, the detection network is trained end-to-end
using the following composite loss function for output tensor $Y$:
\begin{equation}
  \mathcal{L}(Y,\hat{Y}) = \mathcal{L}_{detect}(Y,\hat{Y}) + \mathcal{L}_{group}(Y,\hat{Y}).
\end{equation}

 An L1-regression loss is applied to the detection channels;

\begin{equation}
  \mathcal{L}_{detect}(Y,\hat{Y}) = \sum_{k=1}^{5}\alpha_{ijk}\left|y_{ijk}-\hat{y}_{ijk}\right|,
\end{equation}
where $k$ indexes the landmark channel (the four VB corners and centroid), $i$ and $j$ index
the position in the patch and $\alpha_{ijk}$ is a weighing factor given by 

\begin{equation}
  \alpha_{ijk}= \begin{cases} 
    \frac{N_k}{N_k+P_k} & \mbox{if } \hat{y}_{ijk} \geq T\\
    \frac{P_k}{N_k+P_k} & \mbox{if } \hat{y}_{ijk} < T
  \end{cases}
\end{equation}
where $N_k$ and $P_k$ are the number of pixels in the target detection channel 
respectively less than or greater than some threshold $T$ ($T=0.01$ in this case).

The vector field grouping channels are supervised by an L2-regression loss,
\begin{equation}
  \mathcal{L}_{group}=\sum_{l=1}^4\sum_{b}\sum_{(i,j)\in\mathcal{N}_{bl}}||\mathbf{v}_{ij}^{l}-\mathbf{r}_{ij}^{k}||_2^2.
\end{equation}
Here $l$ indexes each corner type/vector field, $b$ indexes the annotated VBs in the patch
and $\mathcal{N}_{bl}$ is a  neighbourhood surrounding the $l^{th}$ corner of the 
$b^{th}$ VB annotated in that patch. $\mathbf{v}_{ij}^l$ is the value of the output vector
field corrresponding to corner $l$ at location $(i,j)$ and $\mathbf{r}_{ij}^b$ is the
ground truth value of the vector field, ie. the displacement vector from the centroid 
of VB $b$ to location $(i,j)$.

We use heavy augmentation during training including image rotation, rescaling and 
flipping in the coronal plane. The network is trained using an Adam optimizer with 
learning rate $10^{-3}$ with parameters $\mathbf{\beta}=(0.9,0.999)$.

\subsubsection{VB Labelling}

The labelling pipeline requires both the appearance and context networks to be
trained which is done seperately as follows: Firstly a volume is extracted around 
each annotated VB. This is done by tightly fitting a bounding cuboid 
around each detection and then expanding the box by 50\% in each direction to capture
nearby anatomical structures. The resulting volume is the resampled to a size of
$224 \PLH 224 \PLH 16$ voxels (isotropically in the axial and coronal planes but not
in the sagittal plane). These volumes are then given as input to the appearance 
network which attempt to classify the vertebra from S1 to C3 in a 23-way classification
problem. This network is trained on OWS only using a standard cross-entropy loss.

The context network is also trained on OWS. Input height-probability maps are 
contructed such that for a given VB with detected from height $y_a$ to $y_b$,
with centroid at $y_c=\frac{y_a+y_b}{2}$, the height-probability map $P$ has
the same value as the temperature-softmaxed (T=0.1) predictions from the appearance network
from height $y_c - 0.5\times(y_b-y_a)$ to $y_c + 0.5\times(y_b-y_a)$. The context network
is an image-to-image translation network which takes $P$ as input and 
then outputs a refined version of $P$, denoted $P'$. $P'$ is then decoded into a 
discrete series of predictions at the height of each VB detection using a beam 
search. A visual representation of this process is shown in 
Figure~\ref{fig:labelling-pipeline}. As an augmentation during training 
each vertebra detection is dropped from $P$ with probability $0.1$. A loss function
is still applied to the predictions at height of the missing detection, on the basis
that the model should be able to infer the vertebra's level from the surrounding
detections. Both the appearance and context networks are trained using an Adam optimizer with 
learning rate $10^{-3}$ with parameters $\mathbf{\beta}=(0.9,0.999)$.

\subsubsection{Radiological Grading of Degeneration Changes}
The grading network is trained on the radiologist-labelled IVDs 
from the Genodisc dataset from S1-L5 to L1-T12. The model consists of a feature-extracting
backbone that encodes each IVD as a 512-dimensional vector, followed by 8 2-layer MLP 
projection heads, each of which produces predictions for a different grading task.
Each of grading task is highly imbalanced in Genodisc and hence a balanced cross 
entropy loss is used in all cases. 
During training, augmentation is applied by jittering, rotating, flipping and increasing/reducing 
the brightness and contrast of each IVD by $\pm$10\%. In 50\% of the training cases
a random gaussian noise of 10\% is also added. The entire model is trained end-to-end
until 10 consecutive validation epochs do not yield an improvement in averaged balanced accuracy.
This is done using an Adam optimizer with 
learning rate $10^{-4}$ with parameters $\mathbf{\beta}=(0.9,0.999)$. Once this is completed,
the feature-extraction backbone is frozen and the task-specific projections heads are
then trained individually with a learning rate of $10^{-5}$, until 10 consecutive
validation epochs do not yield improved accuracy for that specific grading task.
The results of this training, along with those of the labelling and context networks,
are given in the following Section.

\section{Results}
\label{sec:results}

This section details experimental results from validating the output of
the detection, labelling and grading pipelines on withheld data. 

\subsection{VB Detection and Labelling}

The detection and labelling stages of SpineNetV2 are trained on the training splits
of Genodisc and OWS. Here, we present results of the labelling and grading pipelines
on the test splits from that dataset, as well as on a publically available 
dataset published by Zuki\'c \textit{et al.}~\cite{zukic_robust_2014},
distributed on SpineWeb\footnote{\url{http://spineweb.digitalimaginggroup.ca/}}.
These results are from our initial paper on the method used to detect and
label vertebrae used in SpineNet~\cite{Windsor20b}.

\paragraph{Evaluation:} For the detection stage of the pipeline we report
the precision and recall of the VB detector.
Following~\cite{Lootus13}, we define correct detections to be
when the ground truth vertebra centroid is contained entirely within a single
detected bounding quadrilateral. For the labelling stage, we report the 
identification rate (IDR). To be correctly identified, a vertebra must be
both detected and assigned the correct level by the labelling stage.

\begin{table}[h]
    \centering
    \setlength{\tabcolsep}{3pt}
    {\fontsize{7.55}{9}\selectfont 
    \begin{tabular}{ccccccccc}
    
        \toprule
        Dataset & Scans & Verts & Method & Prec. (\%) & Rec. (\%) & IDR(\%) & IDR$\pm$ 1(\%) & LE (mm)  \\
        \toprule
            \multirow{3}{*}{\shortstack{OWS \\ (Whole \\Spine)}}& & & Windsor\textsuperscript{\textdagger}~\cite{Windsor20}& \textbf{99.4} & \textbf{99.4}  & - & - & \textbf{1.0 $\pm$ 0.9} \\
            & 37 &  888 & Label Baseline & - & - & 86.9 & 93.4 & - \\
            & & & Ours & 99.0 & 98.1 & \textbf{96.5} & \textbf{97.3} & 2.4 $\pm$ 1.3 \\
        \\[-1.5ex]\hline\\[-1.5ex]
        \multirow{3}{*}{\shortstack{Genodisc \\ (Lumbar)}}& & &  Lootus~\cite{Lootus13}& - & - & 86.9 & - & 3.5 $\pm$ 3.3\\
            & 421 & 2947 & Label Baseline & - & - & 90.1  & 97.4 &  -\\
            & & & Ours & \textbf{99.7}  & \textbf{99.7}  & \textbf{98.4} & \textbf{99.7} & \textbf{1.6 $\pm$ 1.1}\\
        \\[-1.5ex]\hline\\[-1.5ex]
        \multirow{3}{*}{\shortstack{Zuki\'c \\ (Lumbar)}}& & & Zuki\'c~\cite{zukic_robust_2014}& 98.7 & 92.9 & - & - & \textbf{1.6 $\pm$ 0.8}\\
        & 17 & 154 & Label Baseline & - & - & 87.0 & 94.3 & -\\
            & & & Ours & \textbf{99.3}  & \textbf{98.7}  & \textbf{90.9} & \textbf{98.7} & 2.0 $\pm$ 1.5\\
        \\[-1.5ex]\hline\\[-1.5ex]
    \end{tabular}
    }
\caption{Performance of the detection and labelling pipeline on the three datasets. 
Our approach is compared with other methods using the same datasets and also a LSTM labelling baseline 
, reported on a per-vertebra level. We also report the percentage of vertebrae 
within one level of their ground truth value (IDR$\pm1$). 
Lootus \cite{Lootus13} is tested on a subset of 291 scans from the Genodisc dataset.
Note, Windsor\textsuperscript{\textdagger}~\cite{Windsor20} requires manual 
initialization by providing the location of the S1 vertebra, 
so is not directly comparable} \label{tab:detection_results}
\end{table}

\paragraph{Baselines:} For each of the three datasets used to assess detection and labelling,
we compare to pre-existing methods reporting results on the same dataset. For OWS,
we use the method outlined in \cite{Windsor20} which detects vertebrae sequentially 
moving up  the spine, starting from S1. It should be noted that this algorithm
requires the location of S1 to be known, and is therefore not directly comparable to 
the our method in that it is only semi-automated. For Genodisc, we compare to
results reported in~\cite{Lootus13}, which detects and labells vertebrae using 
a HOG-template based method in combination with a graphical model. Finally,
for the Zuki\'c dataset, we compare to the results reported in the initial paper.
It should be noted that several other methods have been proposed
to perform vertebra detection and labelling in MRI scan 
(e.g.\ \cite{cai_multi-modal_2016,forsberg_detection_2017,glocker_vertebrae_2013,yang_automatic_2017}).
However, these methods
do not have publically available datasets, and thus direct comparison is not possible.
To motivate the use of a fully-convolutional context network as opposed to a 
more standard recurrent network, we also train a bi-directional LSTM labelling
baseline. This baseline takes features extracted from each vertebra by the appearance
network as a baseline, and outputs a label for each vertebra as output.

\subsection{Radiological Grading}

This section evaluates SpineNetV2's agreement with an expert radiologist
on withheld data from the Genodisc Dataset. Table~\ref{tab:grading_results} 
compares SpineNetV1 and V2
across all V1 tasks (Pfirrmann, Disc Narrowing, Endplate Defects,
Marrow Changes, Spondylolisthesis and Binary Central Canal Stenosis). It
also reports results for new grading tasks added in V2, namely 
4-class central canal stenosis grading, foraminal stenosis grading, and
disc herniation. In all cases the balanced accuracy is reported.

\begin{table}[!htp]\centering
    \scriptsize
    \centering
    \begin{tabular*}{\linewidth}{@{\extracolsep{\fill}}ccccccc}\toprule
    \centering
    Task &Pfirrmann &Disc Narrowing & \multicolumn{2}{c}{Endplate Defect} &\multicolumn{2}{c}{Marrow Change}\\
    & & &Upper &Lower &Upper &Lower \\
    \toprule
    \# Classes &5 &4 &2 &2 &2 &2 \\
    \toprule
    SpineNet V1 &71 &76.1 &82.9 &87.8 & 89.2 &88.4 \\
    SpineNet V2 &70.9 &76.3 &84.9 & 89.6 &88.9 &88.2 \\
    \bottomrule
    \end{tabular*}
    \\
    \begin{tabular*}{\linewidth}{cccccccc}\toprule
     Task &Spondylolisthesis &\multicolumn{2}{c}{Central Canal Stenosis} & \multicolumn{2}{c}{Foraminal Stenosis} & Herniation \\
    & & & &Left &Right & \\
    \toprule
    \# Classes &2 &2 &4 &2 &2 &2 \\
    \toprule
    SpineNet V1 &95.4 &95.8 &- &- &- &- \\
    SpineNet V2 &95 &93.2 &64.9 &84.8 &82.4 &80.4 \\
    \bottomrule
    \end{tabular*}

    \caption{Grading results for withheld data from the Genodisc dataset.
    It should be noted that SpineNetV2 does not have a seperate head for 
    binary central canal stenosis, and instead concatenates together
    predictions from the multiclass central canal stenosis projection head.}
    \label{tab:grading_results}
    \end{table}

As can be seen from Table~\ref{tab:grading_results}, performance of 
SpineNetV2 closely matches that of SpineNetV1 in all tasks. The major
changes are the addition of the 4-class central canal stenosis,
foraminal stenosis grading and disc herniation.

\subsection{Processing Speed}
\label{sec:processing-speed}
For application in real-world scenarios, SpineNetV2 needs to be as fast as possible. Here
we report the inference time of SpineNetV2 with and without a GPU on lumbar and whole spine scans.
We also report the peak memory usage during processing. For a lumbar scan, vertebrae detection and labelling takes
approximately 2 seconds and grading taskes 1 second, whereas for a whole spine scan, detection and labelling 
takes approximately 5 seconds on a GPU. Peak GPU memory usage is 1.6GB for a lumbar scan, 
and 2.5GB for a whole spine scan. On a CPU, lumbar detection and labelling takes 25 seconds,
with grading taking 4 seconds. For a whole spine scans VB detection and labelling takes 
2 minutes on a CPU. This is because the scan is split into multiple patches, each of which
is ingested seperately by the VFR model. This could likely be reduced significantly by changing the 
patch-splitting strategy to use cover larger areas with smaller overlaps. These results are
summarized in Table~\ref{tab:processing-speed}.

 \begin{table}[h]
     \centering
     \begin{tabular*}{\linewidth}{@{\extracolsep{\fill}}cccc}\toprule
     Task Name & Scan Type & \multicolumn{2}{c}{Processing Time (s)} \\
     &&GPU &CPU\\
     \hline
     VB Detection \& Labelling &Lumbar &3 &25\\  
     VB Detection \& Labelling &Whole Spine &5 &120\\  
     IVV Radiological Grading &Lumbar &1 &4\\  
     \end{tabular*}
     \caption{Processing speed of SpineNetV2 on lumbar and whole spine scans.
     Note that, in all cases, SpineNetV2 is much faster than V1 which takes several
     minutes to perform detection and grading on a single lumbar scan using
     a GPU.}

     \label{tab:processing-speed}

\end{table}

\subsection{Example Qualitative Results}

\begin{figure}
    \centering
    \includegraphics[width=\linewidth]{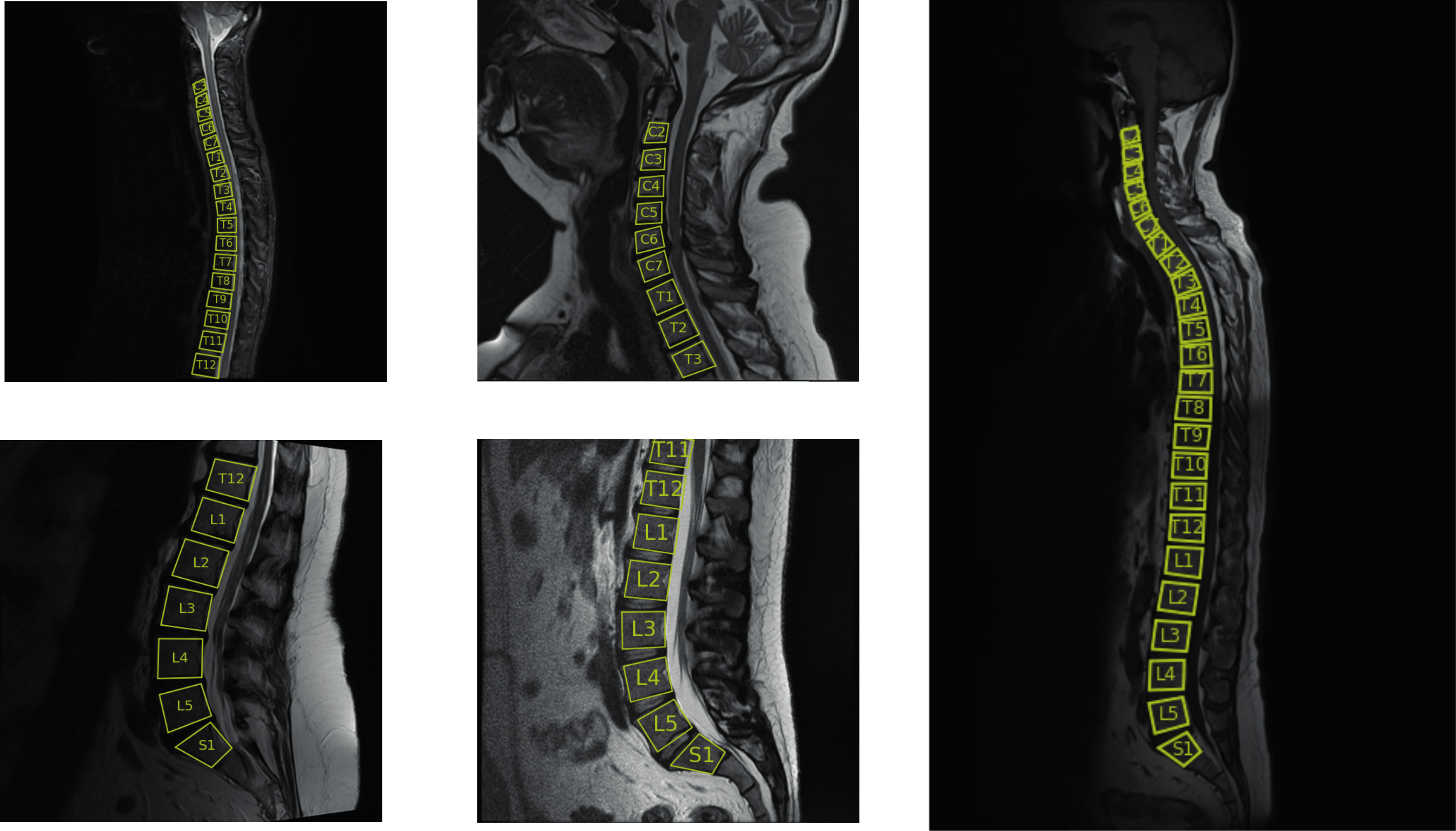}
    \caption{Example VB detections for a variety of fields of view.}
    \label{fig:example-detections}
\end{figure}

Figure~\ref{fig:example-detections} shows example results from the detection pipeline. Example
results from the grading pipeline for healthy and pathological scans
can be seen at~\url{http://zeus.robots.ox.ac.uk/spinenet2/demo.html}.

\subsection{Online Demo}
\begin{figure}[h]
  \centering
  \begin{subfigure}{\linewidth}
    \centering
    \includegraphics[width=.9\linewidth]{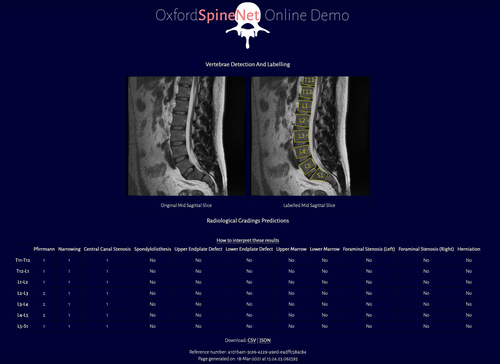}
    \caption{Example healthy spine (\url{radiopedia.org}, rID:33543, Courtesy: A.F. Galliard).}
  \end{subfigure}
  \begin{subfigure}{\linewidth}
    \centering
    \includegraphics[width=.9\linewidth]{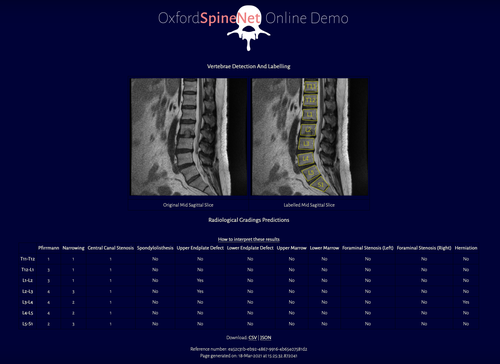}
    \caption{An example spine with degenerative changes (\url{radiopedia.org}, rID:56636, Courtesy: H. Knipe)}
  \end{subfigure}
  \caption{Outputs from the online web demo for two example lumbar MRIs.
  Results such as these can be exported in both CSV and JSON format.}
  \label{fig:webdemo-examples}
\end{figure}

A demo version of the SpineNetV2 software for T2-weighted lumbar MRIs
is available online at~\url{http://zeus.robots.ox.ac.uk/spinenet2/}. This
demo takes as input zipped sagittal slices as DICOMs and outputs a CSV or JSON
containing the grading results, depending on preference.
Cached results for publically available online samples can be found on this 
website. These are shown in Figure~\ref{fig:webdemo-examples}. If you want try the 
demo software on your own scans
please use the contact information provided on the website to be given access. 
Input scans are not stored after processing, however should be anonymised 
prior to submission. Furthermore, please note SpineNetV2 is still a research
tool and should not be used for clinical purposes.

\section{Conclusions}
\label{sec:conclusions}

This technical report describes SpineNetV2, a deep learning framework for
detecting and labelling vertebrae in clinical spinal MR scans and to 
perform radiological grading in T2-weighted scans. We describe the methods
used to in each step of the pipeline, as well as the software architecture
and datasets for training the constituent neural networks. We also give 
experimental results for both VB detection and labelling and radiological 
grading on unseen test data. 

SpineNetV2 is an ongoing project. 
As such we are always looking to validate
the existing functionality of SpineNetV2, as well as extend its functionality 
to new tasks and applications. If you would like to use SpineNetV2 in your
own research, or have any questions, please contact us via email: \texttt{\{rhydian,amir\}@robots.ox.ac.uk}.

\section{Acknowledgements}
We would like to thank our clinical collaborators, without whose sunpport
this project would not be possible; Prof. Jeremy Fairbank, Dr. Jill Urban,
Dr. Sarim Ather, Prof. Ian McCall (in no particular order). We are also
grateful to our funders; Cancer Research UK  via the EPSRC CDT
in Autonomous Intelligent Machines and Systems and EPSRC programme grant Visual
AI (EP/T025872/1).

\FloatBarrier

\bibliographystyle{splncs04}
\bibliography{zotero,vgg_local,vgg_other}

\begin{thebibliography}{10}
\providecommand{\url}[1]{\texttt{#1}}
\providecommand{\urlprefix}{URL }
\providecommand{\doi}[1]{https://doi.org/#1}

\bibitem{cai_multi-modal_2016}
Cai, Y., Landis, M., Laidley, D.T., Kornecki, A., Lum, A., Li, S.: Multi-modal
  vertebrae recognition using {Transformed} {Deep} {Convolution} {Network}.
  Computerized Medical Imaging and Graphics  \textbf{51},  11--19 (2016)

\bibitem{chatfield_return_2014}
Chatfield, K., Simonyan, K., Vedaldi, A., Zisserman, A.: Return of the {Devil}
  in the {Details}: {Delving} {Deep} into {Convolutional} {Nets}. In: {British}
  {Machine} {Vision} {Conference} (2014)

\bibitem{forsberg_detection_2017}
Forsberg, D., Sjöblom, E., Sunshine, J.L.: Detection and {Labeling} of
  {Vertebrae} in {MR} {Images} {Using} {Deep} {Learning} with {Clinical}
  {Annotations} as {Training} {Data}. Journal of Digital Imaging
  \textbf{30}(4),  406--412 (2017)

\bibitem{glocker_vertebrae_2013}
Glocker, B., Zikic, D., Konukoglu, E., Haynor, D.R., Criminisi, A.: Vertebrae
  {Localization} in {Pathological} {Spine} {CT} via {Dense} {Classification}
  from {Sparse} {Annotations}. In: Medical {Image} {Computing} and
  {Computer}-{Assisted} {Intervention}. pp. 262--270. Lecture {Notes} in
  {Computer} {Science} (2013)

\bibitem{guo_calibration_2017}
Guo, C., Pleiss, G., Sun, Y., Weinberger, K.Q.: On {Calibration} of {Modern}
  {Neural} {Networks}. In: International {Conference} on {Machine} {Learning}
  (2017)

\bibitem{Jamaludin16}
Jamaludin, A., Kadir, T., Zisserman, A.: Spinenet: Automatically pinpointing
  classification evidence in spinal mris. In: International Conference on
  Medical Image Computing and Computer Assisted Intervention (2016)

\bibitem{Jamaludin17b}
Jamaludin, A., Kadir, T., Zisserman, A.: Spine{N}et: Automated classification
  and evidence visualization in spinal {MRI}s. Medical Image Analysis
  \textbf{41},  63--73 (2017)

\bibitem{Jamaludin17}
Jamaludin, A., Lootus, M., Kadir, T., Zisserman, A., Urban, J., Batti\'e, M.C.,
  Fairbank, J., McCall, I.: Automation of reading of radiological features from
  magnetic resonance images (mris) of the lumbar spine without human
  intervention is comparable with an expert radiologist. European Spine Journal
   (2017)

\bibitem{Lootus13}
Lootus, M., Kadir, T., Zisserman, A.: Vertebrae detection and labelling in
  lumbar mr images. In: MICCAI Workshop: Computational Methods and Clinical
  Applications for Spine Imaging (2013)

\bibitem{Lootus14}
Lootus, M., Kadir, T., Zisserman, A.: Radiological grading of spinal {MRI}. In:
  MICCAI Workshop: Computational Methods and Clinical Applications for Spine
  Imaging (2014)

\bibitem{palmer_back_2000}
Palmer, K.T., Walsh, K., Bendall, H., Cooper, C., Coggon, D.: Back pain in
  britain: comparison of two prevalence surveys at an interval of 10 years.
  {BMJ}  \textbf{320}(7249),  1577--1578 (2000)

\bibitem{ronneberger_u-net_2015}
Ronneberger, O., Fischer, P., Brox, T.: U-{Net}: {Convolutional} {Networks} for
  {Biomedical} {Image} {Segmentation}. In: Navab, N., Hornegger, J., Wells,
  W.M., Frangi, A.F. (eds.) Medical {Image} {Computing} and
  {Computer}-{Assisted} {Intervention} (2015)

\bibitem{Windsor20}
Windsor, R., Jamaludin, A.: The ladder algorithm: Finding repetitive structures
  in medical images by induction. In: IEEE International Symposium on
  Biomedical Imaging (2020)

\bibitem{Windsor20b}
Windsor, R., Jamaludin, A., Kadir, T., Zisserman, A.: A convolutional approach
  to vertebrae detection and labelling in whole spine mri (2020)

\bibitem{yang_automatic_2017}
Yang, D., Xiong, T., Xu, D., Huang, Q., Liu, D., Zhou, S.K., Xu, Z., Park, J.,
  Chen, M., Tran, T.D., Chin, S.P., Metaxas, D., Comaniciu, D.: Automatic
  {Vertebra} {Labeling} in {Large}-{Scale} {3D} {CT} {Using} {Deep}
  {Image}-to-{Image} {Network} with {Message} {Passing} and {Sparsity}
  {Regularization}. In: Information {Processing} in {Medical} {Imaging}. pp.
  633--644 (2017)

\bibitem{zukic_robust_2014}
Zukić, D., Vlasák, A., Egger, J., Hořínek, D., Nimsky, C., Kolb, A.: Robust
  {Detection} and {Segmentation} for {Diagnosis} of {Vertebral} {Diseases}
  {Using} {Routine} {MR} {Images}. Computer Graphics Forum  \textbf{33}(6),
  190--204 (2014)

\end{thebibliography}

\end{document}